\documentstyle[preprint,aps]{revtex}
\begin{document}
%\draft

\title{Spectroscopy of nanoscopic semiconductor rings}
\author{Axel Lorke, R. Johannes Luyken\cite{AddLuyken}, Alexander O. 
Govorov\cite{AddGovorov}, and J\"org P. Kotthaus}
\address{Sektion Physik und CeNS, LMU M\"unchen, 
Geschwister-Scholl-Platz 1, 80539 M\"unchen, Germany}
\author{J. M. Garcia\cite{AddGarcia} and P. M. Petroff}
\address{Materials Department and QUEST, University of California, Santa 
Barbara, CA 93106}
\date{\today}
\maketitle

\begin{abstract}
Making use of self-assembly techniques, we demonstrate the realization of 
nanoscopic semiconductor quantum rings in which the electronic states are in 
the true quantum limit. We employ two complementary spectroscopic techniques 
to investigate both the ground states and the excitations of these rings. 
Applying a magnetic field perpendicular to the plane of the rings, we find 
that when approximately one flux quantum threads the interior of each ring, a 
change in the ground state from angular momentum $\ell = 0$ to $\ell = -1$ 
takes place. This ground state transition is revealed both by a drastic 
modification of the excitation spectrum and by a change in the magnetic field 
dispersion of the single-electron charging energy.  
\end{abstract}

The fascination of ring-like atomic and quantum structures dates back to 
Kekule's famous proposal of the structure of benzene\cite{Kekule}. 
Particularly interesting are the magnetic properties of such 
non-simply-connected quantum systems, which are related to the possibility to 
trap magnetic flux in their interior. Trapping of a single flux quantum in a 
small molecule such as benzene is impossible with the magnetic fields 
available in today's laboratories. In recent years, however, the availability 
of sub-micron solid-state ring structures has triggered a strong interest in 
the magnetic properties of rings, especially in view of the fact that even in 
the presence of scattering, the many-particle ground state becomes chiral in a 
magnetic field, which leads to so-called ''persistent 
currents''\cite{Buettiker}. The large body of theoretical work on the 
properties of quantum rings\cite{Theorie} is accompanied by a number of ground 
breaking experimental investigations of the magnetic and transport properties 
of rings\cite{Experiment}. These studies have been carried out in the 
mesoscopic range, where scattering still influences the phase coherent 
transport and a large number of quantum states are filled. To the best of our 
knowledge, no spectroscopic data is available on rings in the scatter-free, 
few electron quantum limit. Furthermore, despite a strong theoretical 
interest\cite{Excite,Wendler}, the only data available on the {\em 
excitations} of rings were taken on macroscopic structures\cite{Dahl}. 

Here, we report on the spectroscopy of the ground states and excitations of 
self-assembled, nanoscopic InGaAs quantum rings, occupied with one or two 
electrons each, and subjected to magnetic fields $0 \leq B \leq 12$ T, 
corresponding to 0 -- 1.5 flux quanta threading the interior of the ring. In 
both ground state and excitation spectroscopy we observe characteristic 
changes at magnetic fields around $B = 8$ T which are attributed to the 
development of a magnetic-field-induced chiral ground state.

The quantum rings are fabricated by solid-source molecular-beam epitaxy, using 
the Stranski-Krastanov growth mode, which has now become a well-established 
technique for the fabrication of high-quality, self-assembled semiconductor 
nanostructures\cite{S-K}. Recently, we have reported on a remarkable change in 
morphology when InAs self-assembled dots, grown on GaAs, are covered with a 
thin layer of GaAs and annealed at growth temperature (1 min. at 520 C for the 
present samples) \cite{Garcia,PhysicaB}. Then, the shape of the dots 
drastically changes from that of a lens (roughly 20 nm in diameter, 7 nm in 
height\cite{Leonard}) to one resembling a volcano, with an increased lateral 
size (between 60 and 140 nm in {\it outer} diameter), a reduced height (around 
2 nm) and a well-defined center hole of around 20 nm diameter - see also inset 
in Fig.\ \ref{Fig1}. For the present investigation, these ''self-assembled 
quantum rings'' are embedded in a field-effect transistor (FET) structure, 
which makes it possible to control the number of electrons per ring by 
application of a suitable bias voltage. The electron number can be monitored 
by capacitance-voltage (CV) spectroscopy, which also allows us to derive the 
many-particle ground state energies of the $n$-electron 
system\cite{Drexler,Medeiros}. Furthermore, far-infrared (FIR) transmission 
spectroscopy can be performed simultaneously, to obtain the excitation 
energies of the electronic system\cite{Drexler,Fricke,Miller}. It should be 
pointed out that even though the fabrication of self-assembled rings has been 
demonstrated before\cite{Garcia,PhysicaB}, no conclusive evidence was 
available to show that the surface ring morphology translates into an 
electronic ring structure inside the completed FET. 

The present FET layer sequence starts with a highly Si-doped GaAs back contact 
layer and a 25 nm GaAs spacer, followed by the InAs Stranski-Krastanov layer, 
a 30 nm GaAs cover layer, a 116 nm thick AlAs/GaAs superlattice, and a 4 nm 
GaAs cap. Details of the layer sequence and growth procedure, which are 
essentially identical to those used for quantum dots, can be found in Refs.\ 
\cite{Leonard,Medeiros}. The crucial step for the ring formation is an 
interruption in the growth of the GaAs cover layer after the deposition of a 
nominal thickness $\theta$\cite{Garcia}. On reference samples, which at this 
stage were removed from the growth chamber, we observe ring formation for 
$\theta$ between 1 and 4 nm. The inset in Fig.\ \ref{Fig1} shows an atomic 
force micrograph of a reference sample with $\theta = 2$ nm. 

The electronic spectroscopy was performed at liquid He temperatures on samples 
with an effective area of around 5 mm$^2$, covering approximately $5 \times 
10^8$ rings. The CV traces were taken at low frequencies ($<1$ kHz) using 
standard lock-in technique; the FIR response was recorded by a rapid-scan 
Fourier transform spectrometer. 

The main part of Fig.\ \ref{Fig1} displays CV spectra of samples with 
different $\theta$ but identical layer sequence. These spectra demonstrate the 
strong influence that $\theta$ has, not only on the morphology of the bare 
rings\cite{Garcia}, but also on the electronic properties inside the completed 
FET structure. For $\theta$ = 5 nm, the spectra are indistinguishable from 
those of common quantum dot samples without growth 
interruption\cite{Medeiros,Fricke}: We observe a double peak structure at a 
gate voltage $V_g \approx -1.1$ V, corresponding to the single electron 
charging of the two spin states of the so-called ''$s$-shell'' in the dots. 
Because of inhomogeneous broadening in the present, large-area samples, the 4 
maxima of the $p$-shell\cite{Miller} cannot be well distinguished and have 
merged into a broad plateau around $V_g \approx -0.4$ V. For details on 
quantum dot spectroscopy, see, e.g., Refs. 
\cite{Drexler,Medeiros,Fricke,Miller}. 

For the sample with $\theta$ = 1 nm, two peaks can be clearly distinguished 
around $V_g \approx 0$ V\cite{3rdpeak}. For $\theta = 3$ nm, no individual 
peaks can be identified and only one very broad structure is observed around 
$V_g \approx -0.5$ V. We attribute this to the fact that at the threshold of 
ring formation, the morphology of the Stranski-Krastanov layer is not well 
defined. Nevertheless, this data is of significance, as it is a further 
indication for the structural change that takes place when the growth is 
interrupted. The shift of the first maximum from $-1.2$ V ($\theta = 5$ nm) to 
around 0 V ($\theta = 1$ nm) can be explained by an upward shift of the ground 
state energy caused by the reduced height of the rings compared to the dots. 

In the following, the discussion will focus on the $\theta = 1$ nm ring 
sample. Figure \ref{Fig2}(a) shows the normalized FIR transmission of this 
sample at $V_g = 0.143$ V (upward arrow in Fig.\ \ref{Fig1}) for two different 
magnetic fields $B$, applied perpendicular to the plane of the rings. 
Comparing the carrier density obtained from either CV or FIR spectroscopy with 
the ring density determined by AFM, we find that at this gate voltage, each 
ring is filled with approximately $n_e = 2$ electrons, which shows that (as 
for the dots\cite{Medeiros,Fricke,Miller}) each CV maximum corresponds to the 
filling of one electron per ring.

It should be pointed out here that the FIR measurements, which require a 
signal-to-noise ratio of around 1 part in $10^4$, are extremely challenging 
and at the limit of state-of-the-art FIR spectroscopy. However, from a 
thorough evaluation of a large number of spectra, including subtraction of the 
superimposed signal from the cyclotron resonance in the back contact, we can 
obtain the FIR response as a function of the magnetic field, which is shown in 
Fig.\ \ref{Fig2}(b). 

As indicated by the different symbols, the resonances in Fig.\ \ref{Fig2}(b) 
can be grouped into the following modes: two resonances ($\circ$) which are 
degenerate at $B = 0$ and exhibit orbital Zeeman splitting when a magnetic 
field is applied. A low-lying mode ($\diamond$) which, due to an insufficient 
signal-to-noise ratio at very low energies, can only be detected above 10 meV, 
but extrapolates to $\approx7$ meV at $B = 0$. This mode dies out around $B = 
7$ T, when also the lower $\circ$-mode vanishes and a new mode ($\triangle$) 
appears. The resonances summarized in Fig.\ \ref{Fig2}(b) differ quite 
strongly from those observed in quantum dots\cite{Fricke,Dots}, where in 
general, only two resonances are observed, one of which increases with 
increasing field, whereas the other decreases. On the other hand, Fig.\ 
\ref{Fig2}(b) can be directly compared to the excitation spectrum of quantum 
rings, as calculated, e.g., by Halonen {\it et al.}\cite{Halonen}. Even though 
these calculations were performed for a ring with much larger dimensions, all 
the above experimental features are in good qualitative agreement with the 
calculated energy dispersion. Furthermore, using the effective mass of $m^* 
\approx 0.07$ $m_e$ found in other self-assembled In(Ga)As quantum 
structures\cite{Drexler,Fricke,Miller}, we find that the slopes of the 
$\circ$-modes ($\pm  \frac{1}{2}\hbar \omega_c$, solid lines), the 
$\diamond$-mode ($ \frac{1}{2}\hbar \omega_c$, dotted line), and the 
$\triangle$-mode ($\hbar \omega_c$, dashed line) are all in agreement with the 
calculations. From this we conclude that indeed the morphology seen in Fig.\ 
\ref{Fig1} is preserved when the growth is continued, and that it translates 
into a ring-like electronic structure. As will be discussed in the following, 
the change in the spectrum around $B = 8$ T can then be understood as a direct 
consequence of a magnetic-field-induced change in the ground state. The 
$\times$-mode is not found in the  calculated ring excitations of Ref.\ 
\cite{Halonen}. At present, an explanation for these resonances is still 
missing, and it cannot be ruled out that these resonances originate from the 
presence of a few large quantum dots which have not developed into rings and 
therefore do not change their ground state at $B = 8$ T.

The electronic states in rings can be discussed using a simple model of a 
circular, 1-dimensional wire, bent into a circle of radius $R$. The energy 
levels then follow from the periodic boundary conditions to $E_\ell =  
\frac{\hbar^2}{2m^*}k^2_\ell$ with $k_\ell = \ell \frac{1}{R}$. When a flux 
$\phi = \pi R^2 B$ penetrates the interior of the ring, an additional phase is 
picked up by the electron on its way around the ring, which leads 
to\cite{Theorie} 

\begin{equation}
E_\ell = \frac{\hbar^2}{2m^* R^2}(\ell +  \frac{\phi}{\phi_0})^2 , \qquad \ell 
= 0, \pm1, \pm2, \ldots
\label{eqEn}\end{equation}

with $\phi_0$ being the flux quantum. Thus, with increasing magnetic field, 
the ground state will change from angular momentum $\ell = 0$ to one with 
higher and higher negative $\ell$ (see Fig.\ \ref{Fig3}(a)), a fact intimately 
related to the persistent currents in mesoscopic 
rings\cite{Buettiker,Theorie}. Furthermore, a periodic, Aharonov-Bohm-type 
oscillation in the ground state energy will take place\cite{Theorie}, as shown 
in Fig.\ \ref{Fig3}(a). Obviously, each change in the ground state will in 
turn lead to a pronounced change in the possible transitions, such as the one 
seen in Fig.\ \ref{Fig2}(b) at $B \approx 8$ T. 

For a more quantitative description of the energies and transitions in the 
present rings, we have used a model by Chakraborty {\it et 
al.}\cite{Theorie,Halonen} and calculated the single-particle states in a ring 
potential $U(r) = \frac{1}{2} m^*\omega_0^2 (r - R_0)^2$ (see inset in Fig.\ 
\ref{Fig3}(b)), where $\omega_0$ is the characteristic frequency of the radial 
confinement and $R_0$ is the radius of the ring. These two parameters 
determine the low energy resonance at $B = 0$, estimated at around 7 meV and 
corresponding to an azimuthal excitation $\Delta \ell = \pm1,\ \Delta N = 0$, 
as well as the high energy resonance at 20 meV, corresponding to an excitation 
$\Delta \ell = \pm1,\ \Delta N = 1$, where $N$ is the radial quantum number. 
In this way, both parameters are readily determined to be $R_0 = 14$ nm and 
$\hbar\omega_0 = 12$ meV. The electronic radius of 14 nm is in good agreement 
with the effective radius of the  uncovered rings (i.e. the radius where the 
InGaAs has maximum thickness, see Fig.\ \ref{Fig1}), which is roughly 18 nm. 
Furthermore, we find satisfactory agreement between the calculated and 
measured transitions over the entire range of magnetic fields. Using more 
realistic potentials would certainly improve this agreement, however at the 
cost of introducing additional adjustable parameters. 

The lines in Fig.\ \ref{Fig3}(b) represent the magnetic-field dispersion of 
the ring energy levels, calculated using the model potential $U(r)$ and the 
values for $R_0$ and $\omega_0$ derived above. It can be seen that the 
calculations predict a ground state transition (arrow) from $\ell=0$ to 
$\ell=-1$ at $B \approx 8$ T, in good agreement with the position of the 
change in the FIR resonances. This agreement is somewhat surprising, 
considering that the FIR data was obtained for $n_e = 2$, whereas the model is 
single-particle. From this we conclude that (as in self-assembled InAs 
dots\cite{Fricke,Warburton}) the single particle states are a quite accurate 
basis for the description of the many-particle states and excitations. This 
assumption is supported by the fact that the measured FIR resonance positions 
for $n_e = 1$ are very similar to those for $n_e = 2$, even though the 
signal-to-noise ratio for $n_e = 1$ does not allow us to clearly identify all 
resonances shown in Fig.\ \ref{Fig2}(b). 

There is an additional way to confirm the ground state transition associated 
with the trapping of flux inside the rings. By carefully evaluating the 
position of the lowest capacitance maximum (downward arrow in Fig.\ 
\ref{Fig1}), we have a direct experimental access to the $n_e = 1$ ground 
state energy. The data points in Fig.\ \ref{Fig3}(b) summarize the shift in 
gate voltage of the lowest capacitance peak as a function of the magnetic 
field. As indicated by the arrow, at $B = 8.2$ T, corresponding to a flux of 
$\pi R_0^2 B \approx \phi_0$, a change in the slope can be identified in the 
data. From the comparison with the model calculations, this change in slope is 
identified as the magnetic-field-induced ground state transition from $\ell = 
0$ to $\ell = -1$. 

Note that the shift of the $n_e = 1$ charging peak is given in mV (right hand 
scale in Fig.\ \ref{Fig3}(b)). From a comparison with the calculations (left 
hand scale) we find a voltage-to-energy conversion factor of $f = e \Delta V_g 
/ \Delta E = 1.8$. Using the lever arm model\cite{Leverarm}, which has proved 
to be quite accurate in the case of quantum dots 
\cite{Medeiros,Fricke,Warburton}, the present layer structure gives $f = 7$. 
We attribute this discrepancy to the fact that the slopes at which the $\ell = 
0$ and $\ell = -1$ states intersect strongly depends on the detailed choice of 
the confining potential (as can readily be seen from a comparison between 
Figs.\ \ref{Fig3}(a) and (b)). This sensitivity could in fact provide for a 
useful handle to derive the confining potential from more elaborate model 
calculations.

Converting the separation between the lowest charging peaks in Fig.\ 1 into 
energy, we obtain a Coulomb interaction energy of roughly 20 meV for both 
rings and dots. This similarity is somewhat surprising, given the larger 
lateral size of the rings. The missing central part, which will decrease the 
effective area, may partly be responsible for the large Coulomb interaction in 
the rings. Here, more in-depth theoretical work is desirable, especially in 
view of a possible formation of a rotating Wigner molecule\cite{Wendler} and 
with respect to the fact that experimentally the 1- and 2-electron excitation 
spectra are very similar.

In summary, we have used the striking morphological change that takes place 
when self-assembled InAs quantum dots are partially covered with GaAs to 
fabricate nanoscopic quantum rings with dimensions that bridge the size range 
between mesoscopic and molecular ring structures\cite{Bachtold}. The electron 
states in these rings are not effected by the presence of random scatterers 
and are dominated by quantum effects rather than Coulomb interaction. Using 
two complementary spectroscopic techniques, we have identified a 
magnetic-field-induced transition from a ground state with zero angular 
momentum to a chiral ground state. This transition is a direct consequence of 
the non-simply-connected ring geometry\cite{Moshchalkov} and takes place when 
approximately 1 flux quantum penetrates the effective interior area of the 
rings.

We would like to thank R. J. Warburton, K. Karra\"{\i}, T. Chakraborty, V. 
Gudmundsson, and M. Barranco for helpful and inspiring comments, the latter 
also for making their unpublished work available to us. Financial support from 
QUEST, a NSF Science and Technology Center, and from BMBF, through Grant 01 BM 
623 and a Max Planck research award, is gratefully acknowledged.

%****************************************************************

\begin{figure}
\caption{
Capacitance-voltage (CV) traces for samples with growth interruption after 
different coverages of $\theta = 1$, 3, and 5 nm, respectively. The inset 
displays an atomic force micrograph of self-assembled quantum rings on the 
surface of a reference sample (scan area: 250$\times$250 nm$^2$). 
}
\label{Fig1}
\end{figure}

\begin{figure}
\caption{
(a) Normalized transmission of self-assembled quantum rings (filled with 
$n_e=2$ electrons each) at magnetic fields $B=0$ and $B=10$ T. Minima in the 
transmission (arrows) correspond to electronic excitations. The solid lines 
are a smoothed representation of the data points. Curves are offset for 
clarity. (b) Resonance positions as a function of the magnetic field. 
}
\label{Fig2}
\end{figure}

\begin{figure}
\caption{
(a) Energy levels of an ideal one-dimensional ring as a function of the 
magnetic flux $\phi$ threading the ring area. (b) Calculated energy levels in 
a parabolic wire bent into a circular ring (see inset). The data points (right 
hand scale) give the gate voltage shift of the lowest capacitance maximum.  
}
\label{Fig3}
\end{figure}


\begin{references}

\bibitem[a]{AddLuyken} 
new address: 
Infineon Technologies, Otto Hahn Ring 6, 81739 M\"unchen. 

\bibitem[b]{AddGovorov} 
permanent address: 
Institute of Semiconductor Physics, 630090 Novosibirsk-90, Russia.

\bibitem[c]{AddGarcia} 
new address: 
Instituto de Microelectr\'{o}nica de Madrid, Isaac Newton 8, 28760 Tres 
Cantos, Madrid, Spain. 

\bibitem{Kekule}
A. Kekul\'{e},
Bull. Soc. Chim. France, {\bf 3}, 98 (1865).

\bibitem{Buettiker}
M. B\"uttiker, Y. Imry, and R. Landauer,
Phys. Letters A {\bf 96}, 365 (1983).  

\bibitem{Theorie}
For reviews, see, e.g.,
A. G. Aronov and Yu. V. Sharvin, Rev. Mod. Phys. {\bf 59}, 755 (1987);
T. Chakraborty and P. Pietil\"ainen, 
Phys. Rev. B {\bf 50}, 8460 (1994); 
L. Wendler and V. M. Fomin, 
Phys. Stat. Sol. (b) {\bf 191}, 409 (1995);
and references therein.

\bibitem{Experiment}
L. P. L\'{e}vy {\it et al.},
Phys. Rev. Lett. {\bf 64}, 2074 (1990);
V. Chandrasekhar {\it et al.},
{\it ibid.}, {\bf 67}, 3578 (1991);
D. Mailly, C. Chapelier, and M. Benoit,
{\it ibid.}, {\bf 70}, 2020 (1993);
A. F. Morpurgo {\it et al.},
{\it ibid.}, {\bf 80}, 1050 (1998);
R. Schuster {\it et al.},
Nature {\bf 385}, 417 (1997). 

\bibitem{Excite}
A. Emperador {\it et al.}, 
Phys. Rev. B, {\bf 59}, 15 301 (1999);
I. Magn\'usd\'ottir and V. Gudmundsson, 
http://xxx.lanl.gov/abs/cond-mat/9907216, 
and references therein.

\bibitem{Wendler}
L. Wendler {\it et al.}, 
Phys. Rev. B, {\bf 54}, 4794 (1996);

\bibitem{Dahl}
C. Dahl {\it et al.},
Phys. Rev. B {\bf 48}, 15 480 (1993).

\bibitem{S-K}
L. Landin {\it et al.},
Science {\bf 280}, 262 (1998) and references therein.

\bibitem{Garcia}
J. M. Garc\'{\i}a {\it et al.},
Appl. Phys. Lett. {\bf 71}, 2014 (1997).

\bibitem{PhysicaB}
A. Lorke and R. J. Luyken,
Physica B {\bf 256}, 424 (1998).

\bibitem{Leonard}
D. Leonard {\it et al.},
Appl. Phys. Lett. {\bf 63}, 3203 (1993).

\bibitem{Drexler}
H. Drexler {\it et al.},
Phys. Rev. Lett. {\bf 73}, 2252 (1994).

\bibitem{Medeiros}
G. Medeiros-Ribeiro, D. Leonard, and P. M. Petroff,
Appl. Phys. Lett. {\bf 66}, 1767 (1995).

\bibitem{Fricke}
M. Fricke {\it et al.},
Europhys. Lett. {\bf 36}, 197 (1996). 

\bibitem{Miller}
B. T. Miller {\it et al.},
Phys. Rev. B {\bf 56}, 6764 (1997).

\bibitem{3rdpeak}
On some samples, a third maximum at $V_g \approx 0.2$ V can be identified: 
H. Pettersson {\it et al.}, 
Proc. EP2DS13, to be published in Physica E.

\bibitem{Dots}
C. Sikorski and U. Merkt,
Phys. Rev. Lett. {\bf 62}, 2164 (1989); 
T. Demel {\it et al.},
Phys. Rev. Lett. {\bf 64}, 788 (1990). 

\bibitem{Halonen}
V. Halonen, P. Pietil\"ainen, and T. Chakraborty,
Europhys. Lett. {\bf 33}, 377 (1996). 

\bibitem{Warburton}
R. J. Warburton {\it et al.},
Phys. Rev. B {\bf 58}, 16 221 (1998).

\bibitem{Leverarm}
This model assumes a linear voltage drop between the back contact and the 
front gate, so that the voltage to energy conversion factor is simply given by  
$e d_{In}/d_{top}$, where $d_{In}$ and $d_{top}$ are the distances between the 
back contact and the InGaAs layer / the top gate, respectively. For a 
theoretical assessment of the model's validity, see, e.g. 
O. Heller, Ph. Lelong, and G. Bastard, Physica B {\bf 249}, 271 (1998). 

\bibitem{Bachtold}
Carbon nanotubes are topologically related systems in a similar size range. 
The magnetotransport properties of carbon nanotubes have been studied, e.g., 
by
A. Bachtold {\it et al.},
Nature {\bf 397}, 6721 (1999). 

\bibitem{Moshchalkov}
It is interesting to note that for the discussed ground state transition, the 
non-simply connected geometry is a strict requirement only for {\it 
normal-metallic} systems. In {\it superconducting} systems, on the contrary, 
Aharonov-Bohm-type oscillations can be observed also in a disk geometry 
(V.V. Moshchalkov {\it et al.},
Nature {\bf 373}, 319 (1995).). 
This difference is rooted in the subtleties of the boundary conditions, as 
discussed in 
R. Benoist and W. Zwerger, 
Z. Phys. B {\bf 103}, 377 (1997).

\end{references}
\end{document}